\documentclass[11pt]{article}

\usepackage{graphicx}
\usepackage{amsmath}
\usepackage{amsfonts}
\usepackage{amssymb}
\usepackage{verbatim}

\renewcommand{\arraystretch}{1.3}
\catcode`\@=11
\def\marginnote#1{}

\newcount\hour
\newcount\minute
\newtoks\amorpm
\hour=\time\divide\hour by60
\minute=\time{\multiply\hour by60 \global\advance\minute by-\hour}
\edef\standardtime{{\ifnum\hour<12 \global\amorpm={am}%
        \else\global\amorpm={pm}\advance\hour by-12 \fi
        \ifnum\hour=0 \hour=12 \fi
        \number\hour:\ifnum\minute<10 0\fi\number\minute\the\amorpm}}
\edef\militarytime{\number\hour:\ifnum\minute<10 0\fi\number\minute}

\def\draftlabel#1{{\@bsphack\if@filesw {\let\thepage\relax
      \xdef\@gtempa{\write\@auxout{\string
          \newlabel{#1}{{\@currentlabel}{\thepage}}}}}\@gtempa \if@nobreak
    \ifvmode\nobreak\fi\fi\fi\@esphack} \gdef\@eqnlabel{#1}}
    \def\@eqnlabel{}
\def\@vacuum{}
\def\draftmarginnote#1{\marginpar{\raggedright\scriptsize\tt#1}}

\def\draft{
  \oddsidemargin -.5truein
  \def\@oddfoot{\footnotesize \sl preliminary draft \hfil
    \rm\thepage\hfil\sl\today\quad\militarytime}
  \let\@evenfoot\@oddfoot \overfullrule 3pt
    \let\label=\draftlabel
    \let\marginnote=\draftmarginnote
  \def\@eqnnum{(\theequation)\rlap{\kern\marginparsep\tt\@eqnlabel}%
    \global\let\@eqnlabel\@vacuum}

  }
\makeatletter
\newdimen\normalarrayskip              % skip between lines
\newdimen\minarrayskip                 % minimal skip between lines
\normalarrayskip\baselineskip
\minarrayskip\jot
\newif\ifold             \oldtrue            \def\new{\oldfalse}
\def\arraymode{\ifold\relax\else\displaystyle\fi} % mode of array entries
\def\eqnumphantom{\phantom{(\theequation)}}     % right phantom in eqnarray
\def\@arrayskip{\ifold\baselineskip\z@\lineskip\z@
     \else
     \baselineskip\minarrayskip\lineskip2\minarrayskip\fi}
\def\@arrayclassz{\ifcase \@lastchclass \@acolampacol \or
\@ampacol \or \or \or \@addamp \or
   \@acolampacol \or \@firstampfalse \@acol \fi
\edef\@preamble{\@preamble
  \ifcase \@chnum
     \hfil$\relax\arraymode\@sharp$\hfil
     \or $\relax\arraymode\@sharp$\hfil
     \or \hfil$\relax\arraymode\@sharp$\fi}}
\def\@array[#1]#2{\setbox\@arstrutbox=\hbox{\vrule
     height\arraystretch \ht\strutbox
     depth\arraystretch \dp\strutbox
     width\z@}\@mkpream{#2}\edef\@preamble{\halign
\noexpand\@halignto
\bgroup \tabskip\z@ \@arstrut \@preamble \tabskip\z@ \cr}%
\let\@startpbox\@@startpbox \let\@endpbox\@@endpbox
  \if #1t\vtop \else \if#1b\vbox \else \vcenter \fi\fi
  \bgroup \let\par\relax
  \let\@sharp##\let\protect\relax
  \@arrayskip\@preamble}
\def\eqnarray{\stepcounter{equation}%
              \let\@currentlabel=\theequation
              \global\@eqnswtrue
              \global\@eqcnt\z@
              \tabskip\@centering
              \let\\=\@eqncr

 \halign to \displaywidth\bgroup
    \eqnumphantom\@eqnsel\hskip\@centering
    $\displaystyle \tabskip\z@ {##}$%
    \global\@eqcnt\@ne \hskip 2\arraycolsep
         $\displaystyle\arraymode{##}$\hfil
    \global\@eqcnt\tw@ \hskip 2\arraycolsep
         $\displaystyle\tabskip\z@{##}$\hfil
         \tabskip\@centering
    &{##}\tabskip\z@\cr}
\newfont{\hr}{msbm10}
\newfont{\ams}{msam10}
\hoffset= - 0.9in        % switch off for draft style
\def\beq{\begin{equation}}
\def\eeq{\end{equation}}
\def\ba{\beq\new\begin{array}{c}}
\def\ea{\end{array}\eeq}
\def\be{\ba}
\def\ee{\ea}

\def\d{\partial}
\def\N2{${\cal N}=2$}

\def\1N{${\cal N}=1$}
\def\4N{${\cal N}=4$}
\def\nn{\nonumber}
\def\half{{\textstyle{1\over2}}}

\def\p{\partial}

\newcommand{\rf}[1]{(\ref{#1})}

\def\pha{\phantom.\!\!}

\newdimen\linethick  \linethick=0.4pt
\newdimen\hboxitspace    \hboxitspace=5pt
\newdimen\vboxitspace    \vboxitspace=5pt

\def\fr#1{%
\beq\new
\vcenter{
\hrule height\linethick
          \hbox{\vrule width\linethick
                \kern\hboxitspace
                \vbox{\kern\vboxitspace
                      \hbox{$\begin{array}{c}\displaystyle#1
         \end{array}$}%
                      \kern\vboxitspace}%
                \kern\hboxitspace
                \vrule width\linethick}%
          \hrule height\linethick}%
\eeq}
\renewcommand{\d}{\partial}

\renewcommand{\tt}[1][mer]{\hbox{\tiny{#1}}}

\def\pha{\phantom.\!\!}
\def\vect{\hat}
\def\p{\partial}

\title{{\bf
On Combinatorial Expansions of Conformal Blocks} \vspace{.5cm}}
\author{{\bf A. Marshakov}\thanks{E-mail: \ mars@itep.ru; mars@lpi.ru}\ ,\ \
\date{ } %\\ {\small
{\bf A. Mironov}\thanks{E-mail:
\ mironov@itep.ru; mironov@lpi.ru}
\date{ } \\
{\small {\it Theory Department, Lebedev Physics Institute}
and {\it ITEP, Moscow, Russia}}\\ \\
{\bf A.Morozov}\thanks{E-mail: \ morozov@itep.ru}
\date{ } \\ {\small
{\it ITEP, Moscow, Russia}}}

\begin{document}

\setcounter{footnote}{3}

\setcounter{tocdepth}{3}

\maketitle

\vspace{-8cm}

\begin{center}
\hfill FIAN/TD-14/09\\
\hfill ITEP/TH-29/09
\end{center}

\vspace{5cm}

\begin{abstract}
In a recent paper \cite{agt}
the representation of Nekrasov partition function in terms of
nontrivial two-dimensional conformal field theory
has been suggested. For non-vanishing value of the
deformation parameter $\epsilon=\epsilon_1+\epsilon_2$
the instanton partition function is identified
with a conformal block of Liouville theory with the central charge
$\displaystyle{ c = 1+ 6\epsilon^2/{\epsilon_1\epsilon_2} }$.
If reversed, this observation means that the universal part of conformal blocks,
which is the same for all two-dimensional
conformal theories with non-degenerate Virasoro representations, possesses
a non-trivial decomposition into sum over sets of the Young diagrams,
different from the natural decomposition studied
in conformal field theory. We provide some details about this intriguing
new development in the simplest case of the four-point correlation functions.
\end{abstract}

\tableofcontents

\section{Introduction}

Recent developments in the study of multidimensional nontrivial quantum
field theories are strongly influenced by their resemblance, at least,
in particular issues, to certain features of two-dimensional ($2d$) quantum
field theories. A well-known example of such correspondence is
the $2d$
free field representation of tau-functions, which appear as effective actions
of multidimensional quantum field theories. The universality class of
multidimensional quantum field theories, which can be reformulated in terms of
tau functions of certain integrable systems, include a very interesting
set of supersymmetric gauge theories, where the effective actions are described
by Seiberg-Witten prepotentials \cite{SW}-\cite{DW}.

Nekrasov partition functions \cite{Nek}-\cite{FuMP} generalize
the Seiberg-Witten prepotentials and originally arose from
evaluation of instantonic sums in deformed ${\cal N}=2$
supersymmetric gauge theories. In the language of integrable systems
they look quite similar to combinatorial or character representations
for the tau-functions \cite{chardeco1}, whose quasiclassical counterparts
appeared in description of the
Seiberg-Witten theory in terms of integrable
systems \cite{GKMMM,intSW,intSWbks}.
A non-trivial feature of Nekrasov's functions
from this perspective is their dependence on
extra parameters $\epsilon_1$, $\epsilon_2$ (it is common also
to introduce $\epsilon= \epsilon_1+\epsilon_2$),
which come from the so called $\Omega$-background, providing the IR regularization
of integrals over instanton moduli spaces \cite{MNS,LNS},
but still lacking a clear group-theoretical interpretation.
At $\epsilon=0$, the
Nekrasov partition functions can be represented
in terms of free fermions, or two-dimensional conformal theory
with the central charge $c=1$ \cite{LMN,NO,MN,Tak}, and
overlap partially with particular cases of the
Hurwitz-Kontsevich functions \cite{OP,HK},
where sums over the Young diagrams come from combinatorics
of coverings and appear finally in the well known
form of group characters.
However, the group theory meaning of the deformation to
$\epsilon\neq 0$ remained obscure,
though some formulas were found
that preserve the nice properties of the $\epsilon=0$ case
(see \cite{NY} for the most interesting examples),
and signs of relation to nontrivial conformal
theories with $c\neq 1$ can be seen  already in the relatively
old papers like \cite{Awata}.

Therefore, a recent observation of \cite{agt} which makes
this relation explicit is extremely important.
It reveals deep connections between the Seiberg-Witten theory,
Nekrasov partition functions
and two-dimensional conformal theories in a very transparent,
but somewhat mysterious form.
The claim of \cite{agt}
relates the Liouville CFT model with central charge
$\ \displaystyle{c =1+ 6\epsilon^2/\epsilon_1\epsilon_2}\ $
to a peculiar quiver Nekrasov partition function for the gauge
theories with the powers of $U(2)$ gauge group.
Particular quiver diagrams are associated with
particular (multipoint) conformal blocks, while the gauge group
$SU(2)$ is somehow related to the chiral algebra:
the Virasoro algebra or actually $\widehat{SL(2)}$ \cite{KPZ,Ligr}
in the case of Liouville theory.
Larger gauge groups are presumably related to CFT models with
more free fields and higher $W$-like symmetries \cite{agtplus}.

Unfortunately, while the {\it statements} in \cite{agt}
are very explicit and clear, the underlying
{\it checks} that were actually made are not described
in an equally transparent form.
It took us some effort to reproduce the argument in
the simplest case of the four-point conformal block,
and we present this calculation here, in a hope that
this can help others to easier join this new promising
line of research.

We begin from reminding the classical results
\cite{BPZ,ZZ}
about conformal blocks in $2d$ conformal theory
and then turn to consideration of the observation of \cite{agt}
about their relation to Nekrasov's partition functions.
Exact statement is that conformal block for generic
(i.e.non-degenerate) Verma modules depending on five arbitrary
dimensions $\Delta_i = \alpha_i(\epsilon - \alpha_i)/\epsilon_1\epsilon_2$
is exactly equal to certain linear combination
of Nekrasov's functions, see eq.(\ref{BvsZ}).
This generalizes non-trivially the well-known statement in the case
of the free fields
(see \cite{LMN} and eq.(\ref{BvsZab})),
where dimensions are restricted by two linear constraints:
$\alpha_1+\alpha_2+\alpha_3 + \alpha_4 = \epsilon$
and intermediate-channel $\alpha = \alpha_1+\alpha_2$.
In other words, a switch to the $\alpha$-parametrization of dimensions
appears to convert conformal blocks for Virasoro-Verma modules into
a deformation of character-decomposition formulas,
establishing a one more realization of relation between
the Virasoro and $\widehat{SL(2)}$ algebras.

\section{Conformal block \cite{ZZ}}

\subsection{Conformal block from the operator expansion}

The main ingredients of two-dimensional conformal field theory (2d CFT)
are the states in the Hilbert space, their scalar product,
and the structure constants of the operator algebra.
The states are associated with the
operators $V_{\vect\alpha}(z)$, and the scalar product is determined by the
two-point functions
\be
{\cal K}_{\vect\alpha\vect\beta} = \langle{\vect\alpha}|{\vect\beta}\rangle
\sim \langle V_{\vect\alpha}(0)V_{\vect\beta}(\infty)\rangle,
\label{scapro}
\ee
(where $V_{\vect\beta}(\infty) = \lim_{R\to\infty} R^{2\Delta_{\vect\beta}}V_{\vect\beta}(R)$
with $\Delta_{\vect\beta}$ being dimension of the operator, or quantum number of the corresponding state).
The operator product expansion (OPE)\footnote{
We shall concentrate
in this text, only on the holomorphic (chiral) constituents
(conformal blocks) of the correlation functions.
Physical correlators are bilinear combinations of conformal blocks
and possess global (modular) symmetry.}
is defined as
\be
V_{\vect\alpha}(z)V_{\vect\beta}(z') =
\sum_{\vect\gamma}
\frac{{{\cal C}_{\vect\alpha\vect\beta}^{\vect\gamma}}
\,V_{\vect\gamma}(z')}
{(z-z')^{\Delta_{\vect\alpha}+\Delta_{\vect\beta}
-\Delta_{\vect\gamma}}}
\label{OPE}
\ee
Because of the freedom to
choose the argument of $V_{\vect\gamma}$ at the r.h.s.
the structure constants ${\cal C}$ are actually defined modulo
triangular transformation in the space of $V_{\vect\gamma}$
and their derivatives.
Of course once the choice is made, like in (\ref{OPE}),
$C_{\vect\alpha\vect\beta}^{\vect\gamma}$
are defined unambiguously.
Advantage of the asymmetric choice made in (\ref{OPE})
will be seen in eq.(\ref{4point}) below.

In $2d$ CFT the Hilbert space of states can be described in terms of
representations of
conformal symmetry, generated by the conserved stress-energy tensor $\bar{\partial} T(z)=0$,
i.e. the set of Verma modules,
growing from distinguished (primary or highest-weight) states \cite{BPZ}.
Actually ${\vect\alpha} = \{\alpha,Y\}$ is a multi-index,
where $\alpha$ labels the primary states while $Y$ -- the Young
diagram (integer partition) $Y = \{k_1\geq k_2\geq \ldots \geq k_l>0\}$
labels their descendants, obtained by the action of components of the stress-energy tensor,
or the Virasoro generators
\be
\label{ste}
T(z) = \sum_{n=-\infty}^\infty \frac{L_n}{z^{n+2}}
\ee
on the primary state $V_\alpha = V_{\alpha,\emptyset}$:
\be
V_{\alpha,Y} = L_{-Y} V_\alpha
= L_{-k_l}\ldots L_{-k_2}L_{-k_1} V_\alpha
\label{desc}
\ee
or the highest weight state: $L_k V_\alpha = 0$, for $k>0$, and
$L_0 V_\alpha=\Delta_\alpha V_\alpha$, so that the generators $L_n$ from \rf{ste}
with non-negative $n\geq 0$ do not show up in (\ref{desc}).
Similarly, $[L_{-1},V_\alpha(z)] = \partial V_\alpha(z)$.
Dimension of the descendant is
\be
\Delta_{\vect\alpha} = \Delta_{\alpha,Y}= \Delta_\alpha + |Y|
\ee
where $|Y| = k_1+k_2+\ldots+k_l$ is the size,
i.e. the number of boxes of the Young diagram.
Any combination of Virasoro generators can be brought
to the ordered form (\ref{desc}) with the help of the
Virasoro commutation relation
\be
\left[L_m,L_n\right] = (m-n)L_{m+n} +
\frac{c}{12}n(n^2-1)\delta_{m+n,0}

\ee
where $c$ is the central charge of the theory.

Virasoro symmetry implies that the norms and structure
constants are uniquely defined by those for primary states:
\be
{\cal K}_{\vect\alpha\vect\alpha'} =
{\cal K}_{\alpha,Y;\alpha',Y'}\
=\ K_{\alpha}
\ \delta_{\alpha,\alpha'}\delta_{|Y|,|Y'|}\ Q_{\Delta}(Y,Y'),\\
{\cal C}_{\vect\alpha\vect\alpha'}^{\vect\alpha''} =
%{\cal C}_{\alpha\alpha'}^{\alpha''}(Y,Y';Y'') =
C_{\alpha\alpha'}^{\alpha''}\,
\beta_{\Delta,\Delta'}^{\Delta''}(Y,Y';Y'')
\label{despri}
\ee
and $Q$ and $\beta$ are objects from representation theory
of the Virasoro algebra, entirely independent of the properties
of particular conformal model -- those properties are
fully concentrated in $K$ and $C$.
$Q$ and $\beta$ depend only on dimensions and the central charge
of the model and are important special functions of conformal
field theory.

Practical evaluation of $Q$ and $\beta$ is a straightforward but
tedious problem. By definition, $Q_\Delta(Y,Y')$ is the pair
correlator of two descendants,\footnote{
This correlator behaves as
$\Lambda^{-2\Delta -2|Y'|}$, when the argument $\Lambda$ of
$V_{\Delta }(\Lambda)$ tends to infinity.
One can multiply both sides of the equation by $\Lambda^{2\Delta+2|Y'|}$
and then take the limit, to make the correlator well defined.
Similarly are defined other correlators involving $V(\infty)$, only the correction
factor has to be properly adjusted in each case.}
\be
Q_\Delta(Y,Y') = \langle\Delta | L_Y L_{-Y'}|\Delta\rangle \sim \langle L_{-Y}
V_\Delta(0) L_{-Y'}V_\Delta(\infty)\rangle
\label{Q2p}
\ee
As for $\beta$, the simplest way is to express it through $\gamma$,
a counterpart of $\beta$
for the three-point function\footnote{
Note that because of asymmetry in the definitions (\ref{OPE})
and (\ref{gammadef}),
$\beta$ and $\gamma$ are not symmetric functions even of $Y$ and $Y'$
(since three points in these formulas are fixed
in a special way).
This asymmetry is, however, convenient for the next step:
it simplifies formulas for correlators and conformal blocks. },
\be
\gamma_{\Delta,\Delta',\Delta''}(Y,Y',Y'') =
\sum_{\tilde Y}
\beta_{\Delta,\Delta'}^{\Delta''}(Y,Y';{\tilde Y})Q_{\Delta''}({\tilde Y},Y'')=
\\
= \langle L_{-Y}V_\Delta (0)L_{-Y'}V_{\Delta'}(1)L_{-Y''}V_{\Delta''}(\infty)\rangle
\label{gammadef}
\ee
Moreover, in what
follows for practical calculations we only need the particular cases
$\beta_{\Delta_1\Delta_2}^\Delta(Y) = \beta_{\Delta_1\Delta_2}^\Delta(\emptyset,\emptyset;Y)$ and
\be
\gamma_{\Delta_1\Delta_2\Delta}(Y) = \gamma_{\Delta_1\Delta_2\Delta}(\emptyset,\emptyset;Y) =
\sum_{\tilde Y}
\beta_{\Delta,\Delta'}^{\Delta}({\tilde Y})Q_{\Delta}({\tilde Y},Y)
\label{gdef}
\ee
Given (\ref{OPE}), the four-point function can be written as follows:
\be
\langle V_{\vect\alpha_1}(z_1) V_{\vect\alpha_2}(z_2)
V_{\vect\alpha_3}(z_3) V_{\vect\alpha_4}(z_4)\rangle \ \
\stackrel{(\ref{OPE})}{=}\!\!\!\! = \\
\sum_{\vect\alpha_{12},\vect\alpha_{34}}\!
\frac{{\cal C}_{\vect\alpha_1\vect\alpha_2}^{\vect\alpha_{\!(12)}}
{\cal C}_{\vect\alpha_3\vect\alpha_4}^{\vect\alpha_{\!(34)}}}
{(z_1-z_2)^{\Delta_{\vect\alpha_1}+\Delta_{\vect\alpha_2}-\Delta_{\vect\alpha_{12}}}
(z_3-z_4)^{\Delta_{\vect\alpha_3}+\Delta_{\vect\alpha_4}-\Delta_{\vect\alpha_{34}}}}
\ \langle V_{\vect\alpha_{\!(12)}}(z_2) V_{\vect\alpha_{\!(34)}}(z_4)\rangle
\label{4point}
\ee
If we now put $z_2=\infty$ and $z_4=0$, then the two-point
function at the r.h.s. turns into the scalar product (\ref{Q2p})
-- this is what justifies the
asymmetric definition in (\ref{OPE}).
It remains to put $z_1=1$, $z_3=x$,
take {\it primaries} for the four original operators,
i.e. put $\vect\alpha_i = \{\alpha_i,\emptyset\}$ for $i=1,2,3,4$,
and make use of the factorization formulas (\ref{despri}).
Finally we pick up a contribution of particular
intermediate {\it channel} $\alpha_{12}=\alpha_{34}=\alpha$
in the sum at the r.h.s. of (\ref{4point})
-- this gives the {\it definition} of the 4-point conformal block:
\be
\boxed{
{\cal F}_{\Delta}(\Delta_1,\Delta_2;\Delta_3,\Delta_4|x) \equiv
x^\sigma \left(C_{\alpha_1,\alpha_2}^\alpha {\cal K}_\alpha
C_{\alpha_3,\alpha_4}^\alpha\right)
{\cal B}_{\Delta}(\Delta_1,\Delta_2;\Delta_3,\Delta_4|x)
}
\label{cobfac}
\ee
where $\sigma$ is a simple combination of dimensions which we do
not need in his paper, and
${\cal B}$ is a pure representation-theory quantity:
\fr{
{\cal B}_{\Delta}(\Delta_1,\Delta_2;\Delta_3,\Delta_4|x) =
\sum_k x^k{\cal B}^{(k)}_\Delta=
\sum_{|Y|=|Y'|} x^{|Y|} {\cal B}_{Y,Y'} =
\\
 =
\sum_{|Y|=|Y'|} x^{|Y|}\beta_{\Delta_1\Delta_2}^\Delta(Y)
Q_\Delta(Y,Y')\beta_{\Delta_3\Delta_4}^\Delta(Y')
\\
=\sum_{|Y|=|Y'|} x^{|Y|}\gamma_{\Delta_1\Delta_2\Delta}(Y)
Q^{-1}_\Delta(Y,Y')
\gamma_{\Delta_3\Delta_4\Delta}(Y')
\label{calBdef}}
Note that even when the channel is fixed, the sums survive over
Virasoro descendants $V_{\vect\alpha}$ of the primary field $V_\alpha$,
moreover these descendants can be different in $\vect\alpha_{\!(12)}
= \{\vect\alpha,Y\}$ and $\vect\alpha_{\!(34)} = \{\vect\alpha,Y'\}$,
therefore (\ref{calBdef}) is actually a {\it double} sum
over Young diagrams.
We emphasize once again that $\beta$ and $\gamma$ are highly asymmetric
functions of the three dimensions in their arguments.

Similarly one can define multi-point conformal blocks.
For this one needs to know a more general quantity than $\gamma(Y)$
from the Virasoro-group theory: $\gamma(Y_1,Y_2,\emptyset)$
which depends on two different Young diagrams.
{\it Two} reduces to {\it one} for the four-point functions.
We leave generic conformal blocks beyond consideration in this short text
(but note that AGT conjecture \cite{agt} is made in full generality
and involves multi-point conformal blocks).

Many considerations in CFT involve Verma modules with null-vectors,
when relevant representations of the Virasoro algebra
are actually factors of Verma modules --
then formula like (\ref{calBdef}) involve sums over these factors
only and modified quantities appear there instead of $K$ and $\beta$.
As in \cite{agt},
we consider here only the generic case of non-degenerate Verma modules.

\subsection{The 4-point conformal block for arbitrary Verma module
\label{4pcb}}

In order to evaluate ${\cal B}$ from (\ref{calBdef}) one needs to
know just two ingredients: the triple vertex $\gamma$ and
Shapovalov form $Q$ for Virasoro algebra.
They are described by the simple formulas
(in a slightly
abbreviated notation for $Q_\Delta(Y,Y')=Q_\Delta([k_1k_2\ldots],[k'_1k'_2\ldots])$):
\be
 Q_\Delta(Y,Y') =
\label{table}
\ee
\be
\begin{array}{|c||c||c||c|c||c|c|c||c|}
\hline
Y/Y' & \emptyset & [1] & [2] & [11] & [3] & [21] & [111] &\ldots\\
\hline\hline
\emptyset & 1 &&&&&&&\\
\hline\hline
[1] &&2\Delta&&&&&&\\
\hline\hline
[2] &&&\half(8\Delta + c)&6\Delta&&&&\\
\hline
[11] &&&6\Delta&4\Delta(1+2\Delta)&&&&\\
\hline\hline
[3] &&&&&6\Delta+2c&2(8\Delta+c)&24\Delta&\\
\hline
[21] &&&&&2(8\Delta+c)&8\Delta^2+(34+c)\Delta + 2c&36\Delta(\Delta+1)&\\
\hline
[111] &&&&&24\Delta&36\Delta(\Delta+1)&24\Delta(\Delta+1)(2\Delta+1)&\\
\hline\hline
\ldots &&&&&&&&\\
\hline
\end{array}
\nn
\ee
and for arbitrary three primaries
\be
\boxed{
\gamma_{\Delta_1,\Delta_2;\Delta}(Y)
= \prod_i \left( \Delta + k_i\Delta_1 - \Delta_2
+  \sum_{j< i} k_j\right) \sim
\langle L_{-Y}V_\Delta(0) V_{\Delta_1}(1)V_{\Delta_2}(\infty)\rangle
}
\label{gammaprod}
\ee
In particular,
\be
\gamma_{\Delta_1,\Delta_2;\Delta}[1]=\Delta+\Delta_1-\Delta_2,\nn\\ \nn\\
\gamma_{\Delta_1,\Delta_2;\Delta}[2]=\Delta+2\Delta_1-\Delta_2,\nn\\
\gamma_{\Delta_1,\Delta_2;\Delta}[11]=
(\Delta+\Delta_1-\Delta_2)(\Delta+\Delta_1-\Delta_2+1),\nn\\ \nn\\
\gamma_{\Delta_1,\Delta_2;\Delta}[3]=\Delta+3\Delta_1-\Delta_2,\nn\\
\gamma_{\Delta_1,\Delta_2;\Delta}[21]=
(\Delta+2\Delta_1-\Delta_2)(\Delta+\Delta_1-\Delta_2+2),\nn\\
\gamma_{\Delta_1,\Delta_2;\Delta}[111]=
(\Delta+\Delta_1-\Delta_2)(\Delta+\Delta_1-\Delta_2+1)
(\Delta+\Delta_1-\Delta_2+2),\nn\\
\ldots\nn \\
\gamma_{\Delta_1,\Delta_2;\Delta}[n]=\Delta+n\Delta_1-\Delta_2,\nn\\
\ldots\nn\\
\gamma_{\Delta_1,\Delta_2;\Delta}[1^n]=
(\Delta+\Delta_1-\Delta_2)(\Delta+\Delta_1-\Delta_2+1)\ldots
(\Delta+\Delta_1-\Delta_2+n-1)
\label{gammaexa}
\ee
Substituting these explicit formulas
into (\ref{calBdef}), we obtain:
\be
{\cal B}^{(0)}=1\nn
\ee
\be
{\cal B}_\Delta^{(1)} =
{(\Delta+\Delta_1 -\Delta_2   )(\Delta+\Delta_3-\Delta_4)
\over 2\Delta}
\label{B1}
\ee
\be
{\cal B}_\Delta^{(2)}=
{(\Delta+\Delta_1 -\Delta_2   )(\Delta+\Delta_1 -\Delta_2   +1)
(\Delta+\Delta_3-\Delta_4)
(\Delta+\Delta_3-\Delta_4+1)\over 4\Delta(2\Delta+1)}+\\
+{\left[(\Delta_2   +\Delta_1 )(2\Delta+1)+\Delta(\Delta-1)
-3(\Delta_2   -\Delta_1 )^2\right]
\left[(\Delta_3+\Delta_4)(2\Delta+1)+\Delta(\Delta-1)
-3(\Delta_3-\Delta_4)^2\right]
\over 2(2\Delta+1)\Big(2\Delta(8\Delta-5) + (2\Delta+1)c\Big)}
\label{B2}
\ee
\be
{\cal B}_\Delta^{(3)}=
{1\over 2\Delta (3 \Delta ^2+c \Delta -7 \Delta +2+c)}\left[
(\Delta +3 \Delta_1  -\Delta_2    )(\Delta ^2+3 \Delta +2)
(\Delta +3 \Delta_3 -\Delta_4 ) -\phantom{1\over 2}\right.\\
-2(\Delta +3 \Delta_1  -\Delta_2    )(\Delta +1)
 (\Delta +2 \Delta_3 -\Delta_4 )(\Delta +\Delta_3 -\Delta_4 +2) +\\
 + (\Delta +3 \Delta_1  -\Delta_2    )(\Delta +\Delta_3 -\Delta_4 )
 (\Delta +\Delta_3 -\Delta_4 +1) (\Delta +\Delta_3 -\Delta_4 +2) -\\
 -2(\Delta +2 \Delta_1  -\Delta_2    ) (\Delta +\Delta_1  -\Delta_2    +2)
 (\Delta +1)(\Delta +3 \Delta_3 -\Delta_4 ) + \\
 +(\Delta +\Delta_1  -\Delta_2    ) (\Delta +\Delta_1  -\Delta_2    +1)
 (\Delta +\Delta_1  -\Delta_2    +2)(\Delta +3 \Delta_3 -\Delta_4 )+
\label{B3}
\ee
$$\hspace{-1cm}+2\frac{(\Delta +2 \Delta_1  -\Delta_2    ) (\Delta +\Delta_1  -\Delta_2    +2)
 (6 \Delta ^3+9 \Delta ^2-9 \Delta +2 c \Delta ^2+3 c \Delta +c)
 (\Delta +2 \Delta_3 -\Delta_4 )(\Delta +\Delta_3 -\Delta_4 +2)}
 {16\Delta^2+2(c-5)\Delta+c}
 $$ $$ \hspace{-1cm}-\frac{(\Delta +2 \Delta_1  -\Delta_2    )(\Delta +\Delta_1  -\Delta_2    +2)
 (9 \Delta ^2-7 \Delta +3 c \Delta +c)(\Delta +\Delta_3 -\Delta_4 )
 (\Delta +\Delta_3 -\Delta_4 +1)(\Delta +\Delta_3 -\Delta_4 +2)}
 {16\Delta^2+2(c-5)\Delta+c}
 $$ $$\hspace{-1cm}-\frac{(\Delta +\Delta_1  -\Delta_2    ) (\Delta +\Delta_1  -\Delta_2    +1)
 (\Delta +\Delta_1  -\Delta_2    +2)(9 \Delta ^2-7 \Delta +3 c \Delta +c)
 (\Delta +2 \Delta_3 -\Delta_4 )(\Delta +\Delta_3 -\Delta_4 +2)}
 {16\Delta^2+2(c-5)\Delta+c}
 $$ $$ \hspace{-1cm}+ (\Delta +\Delta_1  -\Delta_2    ) (\Delta +\Delta_1  -\Delta_2    +1)
  (\Delta +\Delta_1  -\Delta_2    +2)
  (\Delta +\Delta_3 -\Delta_4 )(\Delta +\Delta_3 -\Delta_4 +1)
  (\Delta +\Delta_3 -\Delta_4 +2)\times
  $$ $$
  \left.
  \times \frac{
  (24 \Delta ^2-26 \Delta +11 c \Delta +8 c+c^2)
  }
  {12\Big(16\Delta^2+2(c-5)\Delta+c\Big)}\right]
$$
\be
\ldots
\nn
\ee
Eqs.(\ref{B1}) and (\ref{B2})
are exactly the well-known formulas from \cite{ZZ}
(with misprints corrected).

\subsection{Comments on the derivation of $Q$ and $\gamma$}

In this section we follow the standard basic of conformal theory
\cite{BPZ,ZZ}.
Evaluation of the Shapovalov form $Q_\Delta$ is absolutely straightforward.
By definition \rf{Q2p} it is
non-vanishing only for $|Y|=|Y'|$.
The first few
elements of the matrix $Q_\Delta(Y,Y')$ for \rf{table} can be easily calculated
\be
Q_\Delta(\emptyset,\emptyset) = \langle\Delta|\Delta\rangle = 1;
\\
Q_\Delta([1],[1]) = \langle\Delta| L_1 L_{-1}|\Delta\rangle =
\langle\Delta| L_{-1} L_{1} + 2 L_0|\Delta\rangle =
(0+2\Delta)\langle\Delta|\Delta\rangle = 2\Delta
\label{Q1,1}
\ee
for $|Y|=0$ and $|Y|=1$. Next,
\be
\pha[2],[2]:\hfill L_{2}L_{-2} = L_{-2}L_{2} +
4L_0 + \frac{c}{2}
\rightarrow 4\Delta + c/2 = \frac{1}{2}(8\Delta + c),  \\
\pha[2],[11]:\hfill  L_2L_{-1}^2 = (L_{-1}L_2+3L_1)L_{-1}
\rightarrow 3L_1L_{-1} = 3L_{-1}L_1 + 6L_0 \rightarrow 6\Delta,\\
\pha[11],[11]:\hfill  L_1^2L_{-1}^2 = L_1(L_{-1}L_1 + 2L_0)L_{-1}
= L_1L_{-1}(L_{-1}L_1 + 2L_0) + 2L_1(L_{-1}L_0+L_{-1})\rightarrow\\
\rightarrow 0 + \big((2+2)\Delta + 2\big)L_1L_{-1}
\stackrel{(\ref{Q1,1})}{\longrightarrow} 4\Delta(1+2\Delta)
\label{Q2,2}
\ee
for $|Y|=2$, and
\be
\pha[3],[3]:\hfill  L_3L_{-3} \rightarrow 6\Delta + 2c, \nn \\
\pha[3],[21]: \hfill  L_3L_{-1}L_{-2} \rightarrow 4L_2L_{-2}
\stackrel{(\ref{Q2,2})}{\longrightarrow} 2(8\Delta +c),\nn\\
\pha[3],[111]: \hfill  L_3L_{-1}^3 \rightarrow 4L_2L_{-1}^2
\stackrel{(\ref{Q2,2})}{\longrightarrow} 24\Delta, \nn \\
\pha[21],[21]: \hfill  L_2L_1L_{-1}L_{-2} = L_2(L_{-1}L_1+2L_0)L_{-2}
\rightarrow 3L_2L_{-1}^2 + 2L_2L_{-2}(\Delta+2)\rightarrow \nn\\
\ \ \ \ \ \stackrel{(\ref{Q2,2})}{\longrightarrow}
18\Delta + (8\Delta+c)(\Delta+2) = 8\Delta^2 + (34+c)\Delta + 2c,\nn\\
\pha[21],[111]: \hfill  L_2L_1L_{-1}^3 = L_2(L_{-1}L_1 + 2L_0)L_{-1}^2
= L_2L_{-1}(L_{-1}L_1 + 2L_0)L_{-1} + 2L_2(L_{-1}L_0+L_{-1})L_{-1}
\rightarrow \nn \\
\ \ \ \ \ \rightarrow 2L_2L_{-1}^2\big(\Delta+(\Delta+1)+(\Delta+1)+1\big)
\stackrel{(\ref{Q2,2})}{\longrightarrow}
36\Delta(\Delta+1), \nn \ee
\be
\pha[111],[111]: \hfill  L_1^3L_{-1}^3 = L_1^2(L_{-1}L_1 + 2L_0)L_{-1}^2
= L_1^2L_{-1}(L_{-1}L_{1}+2L_0)L_{-1} + 2L_1^2(L_{-1}L_0+L_{-1})L_{-1}
\rightarrow \nn \\
\ \ \ \ \ \rightarrow 2L_1^2L_{-1}^2\big(\Delta+(\Delta+1) + (\Delta+1) + 1\big)
\stackrel{(\ref{Q2,2})}{\longrightarrow}
24\Delta(1+\Delta)(1+2\Delta)
\ee
Generally, for example
\be
Q_\Delta([n],[n]) = \langle\Delta| L_n L_{-n}|\Delta\rangle =
\langle\Delta| L_{-n} L_{n} + \frac{c}{12}n(n^2-1)
+ 2n L_0|\Delta\rangle
= \frac{c}{12}n(n^2-1) + 2n\Delta, \\
Q_\Delta([n],[n-1,1])
= \langle\Delta| L_n L_{-1} L_{-(n-1)}\Delta\rangle =
(n+1)\langle\Delta| L_{n-1} L_{-(n-1)} |\Delta\rangle = \\
= (n+1)Q([n-1],[n-1]|\Delta)
= \frac{c}{12}(n+1)n(n-1)(n-2) + 2(n^2-1)\Delta
\ee
and so on, up to
\be
Q_\Delta([1^n],[1^n]) = \langle\Delta| L_1^n L_{-1}^n|\Delta\rangle
= n(2\Delta + n-1)
\langle\Delta| L_1^{n-1} L_{-1}^{n-1}|\Delta\rangle
= n!\frac{\Gamma(2\Delta+n)}{\Gamma(2\Delta)}
\ee
The last formula follows from two recurrent relations:
\be
L_0L_{-1}^n = L_{-1}L_0L_{-1}^{n-1} + L_{-1}^n = \ldots
= L_{-1}^n(L_0+n) = L_{-1}^n(\Delta + n)
\label{0-1n}
\ee
and
\be
L_1L_{-1}^n = L_{-1}L_1L_{-1}^{n-1} + 2L_0L_{-1}^{n-1}
= \ldots
\ \stackrel{(\ref{0-1n})}{=}\ L_{-1}^nL_1 + \nn \\
+ 2L_{-1}^{n-1}\Big(
(\Delta+n-1) + (\Delta+n-2) + \ldots + \Delta\Big)
\rightarrow 0 + n(2\Delta+n-1)L_{-1}^{n-1}
\ee
The nice factorized formula (\ref{gammaprod}) to the three-point function $\gamma$
can be derived \cite{Be} using the general theory \cite{BPZ,ZZ}.
The simplest way to obtain (\ref{gammaprod})  for generic $Y$,
is to use explicit expression
for the action of the Virasoro generators on the primary fields
\be
[L_n,V_\Delta (z)]=\left(z^{n+1}{d\over dz}+(n+1)\,z^n\Delta\right)V_\Delta (z)
\ee
and the manifest expression for the holomorphic part of the
3-point correlator of the primary fields
\be
\langle V_{\Delta_1}(z_1)V_{\Delta_2}(z_2)V_{\Delta_3}(z_3)\rangle =
{C_{\Delta_1,\Delta_2,\Delta_3}\over
(z_1-z_2)^{\Delta_1+\Delta_2-\Delta_3}(z_1-z_3)^{\Delta_1+\Delta_3-\Delta_2}
(z_2-z_3)^{\Delta_2+\Delta_3-\Delta_1}}
\ee
taken at points $z_1=0$, $z_2=z$, $z_3=\infty$ ($z$ is put 1 after calculating all
derivatives).

The simplest example is
\be
\langle V_{\Delta_1}(z_1)  V_{\Delta_2}(z_2) L_{-1} V_\Delta(z) \rangle
= \frac{\partial}{\partial z}
\langle V_{\Delta_1}(z_1)  V_{\Delta_2}(z_2) V_\Delta(z) \rangle = \\
= \frac{\partial}{\partial z}
\frac{C_{\Delta_1,\Delta_2\Delta_3}}{(z-z_1)^{\Delta+\Delta_1-\Delta_2}
(z-z_2)^{\Delta+\Delta_2-\Delta_1} (z_1-z_2)^{\Delta_1+\Delta_2-\Delta}}
\ee
Only derivative of the first factor with $(z-z_1)$ contributes when we
put $z_2=\infty$, and this explains why
$\gamma(\Delta_1,\Delta_2,\Delta) = \Delta+\Delta_1-\Delta_2$.

\subsection{Free fields}

The simplest and most well known example of conformal theory with
{\it arbitrary} central charge
$c = 1+6Q^2$ is the theory of free massless field $\phi$
with background charge $Q$,
considered to be a generic real or even complex number.
The primaries $V_\alpha = \ :e^{\alpha\phi}:\ $ in this theory
have conformal dimensions
\be
\Delta_\alpha = \alpha(Q-\alpha)
\label{ffdims}
\ee
The OPE in this theory is very simple:
\be
:e^{\alpha_1\phi(z_1)}:\ :e^{\alpha_2\phi(z_2)}:\
= \frac{
:e^{\alpha_1\phi(z_1)+\alpha_2\phi(z_2)}:}
{(z_1-z_2)^{2\alpha_1\alpha_2}} =   \\
= z_{12}^{-2\alpha_1\alpha_2}:\left(1 + z_{12}\alpha_1\p\phi(z_2) +
z_{12}^2 \frac{\alpha_1^2}{2}\Big(\p\phi(z_2)\Big)^2
+ z_{12}^2 \frac{\alpha_1}{2}\p^2\phi(z_2) + \ldots\right)
e^{(\alpha_1+\alpha_2)\phi(z_2)}:\ =  \\
=z_{12}^{-2\alpha_1\alpha_2}\Big(1 +
z_{12}\beta_{free}{[1]}\,L_{-1} +
z_{12}^2(\beta_{free}{[2]}\,L_{-2}
+
\beta_{free}{[11]}\,L_{-1}^2) + \ldots\Big)
\! :e^{(\alpha_1+\alpha_2)\phi(z_2)}:
\label{ffope}
\ee
and leads to the following selection rule for the four-point conformal block
\be
\alpha\equiv\alpha_{(12)}=\alpha_1+\alpha_2 = Q-\alpha_3-\alpha_4,
\label{ffcons}
\ee
It means that in the intermediate channel there is a single primary operator
with the conformal dimension
\be
\Delta = \alpha(Q-\alpha) = (\alpha_1+\alpha_2)(Q-\alpha_1-\alpha_2)
= (Q-\alpha_3-\alpha_4)(\alpha_3+\alpha_4) =
(\alpha_1+\alpha_2)(\alpha_3+\alpha_4)
\label{Deltafree}
\ee
The four-point conformal block in this theory just
equals
\be
\langle
e^{\alpha_1\phi(z_1)}e^{\alpha_2\phi(z_2)}
e^{\alpha_3\phi(z_3)}e^{\alpha_4\phi(z_4)}\rangle
= \prod_{i<j} z_{ij}^{-2\alpha_i\alpha_j}
\label{cbQ0}
\ee
with $\sum_i \alpha_i = Q$.
Putting $z_1=1$, $z_2=\infty$, $z_3=x$, $z_4=0$
as requested in (\ref{cobfac}),
then the r.h.s. turns into
\be
\label{frebl}
{\cal B}_{free}(x) =
x^{-2\alpha_3\alpha_4}(1-x)^{-2\alpha_1\alpha_3} =
x^{-2\alpha_3\alpha_4}
\sum_{k=0}^\infty x^k \frac{\Gamma(k+2\alpha_1\alpha_3)}
{k!\,\Gamma(2\alpha_1\alpha_3)} =  \\
= x^{-2\alpha_3\alpha_4}\Big(1 + 2\alpha_1\alpha_3 x
+ \alpha_1\alpha_3(2\alpha_1\alpha_3+1)x^2 +
\frac{2\alpha_1\alpha_3(\alpha_1\alpha_3+1)(2\alpha_1\alpha_3+1)}{3}\,x^3
+ \ldots\Big)
\ee
In particular, the coefficient in the first term of expansion
in brackets is $2\alpha_1\alpha_3$,
what should be compared with (\ref{B1}) which in this
case is equal to
\be
{\cal B}_{free}^{(1)}\!\stackrel{(\ref{B1})}{=}
{(\Delta+\Delta_1 -\Delta_2   )(\Delta+\Delta_3-\Delta_4)
\over 2\Delta} =
\frac{2\alpha_1(Q-\alpha_1-\alpha_2)\cdot 2\alpha_3(Q-\alpha_3-\alpha_4)}
{2(\alpha_1+\alpha_2)(\alpha_3+\alpha_4)}
\ \stackrel{(\ref{ffcons})}{=}\
2\alpha_1\alpha_3
\label{B1free}
\ee
In a similar way one can check that the next two coefficients
coincide with the values of ${\cal B}^{(2)}$ and
${\cal B}^{(3)}$ obtained when the free-field dimensions
and (\ref{Deltafree})
are substituted into (\ref{B2}) and (\ref{B3}):
\be
{\cal B}_{free}^{(2)}\!\stackrel{(\ref{B2})}{=}
\frac{2\alpha_1\alpha_3(2\alpha_1\alpha_3+1)}{2!},   \\
{\cal B}_{free}^{(3)}\!\stackrel{(\ref{B3})}{=}
\frac{2\alpha_1\alpha_3(2\alpha_1\alpha_3+1)(2\alpha_1\alpha_3+2)}{3!}
\label{B23free}
\ee
Note that they are independent of $Q$, despite particular
dimensions $\Delta_{\alpha_a}$ are $Q$-dependent.
Naturally, one expects that if explicit expressions like
(\ref{B1})-(\ref{B3}) are found for higher terms of the $x$-expansion,
then substitution of (\ref{Deltafree}) would give
\be
{\cal B}_{free}^{(k)}=
\frac{\Gamma(k+2\alpha_1\alpha_3)}
{k!\,\Gamma(2\alpha_1\alpha_3)}
\label{Bmfree}
\ee
To summarize, the representation theory formula (\ref{calBdef})
provides an amusing
expansion of the free field correlator \rf{frebl} into
a bilinear sum over Young diagrams of equal sizes.

\subsection{Character decomposition
\label{ss:comb}}

Expression \rf{frebl} (for further convinience in this section we denote
$\sqrt{2}\alpha_1=m_1$, $\sqrt{2}\alpha_3=m_2$; we shall see below that they play the role
of the rescaled\footnote{
This rescaling is just an artefact of inconvenient normalizations, chosen originally
in \cite{BPZ,ZZ}
and can be absorbed by renormalization of the scalar field $\phi(z)=\sqrt{2}\varphi(z)$
(cf. e.g.
normalizations in \cite{fref} and \cite{gmmos}).}
mass parameters $\mu$ in 4d theory)
has an obvious alternative expansion into the single sum over partitions (see
e.g. \cite{McD})
\be
\label{c1expn}
(1-x)^{-m_1m_2} = \sum_k {x^k\over k!}{\Gamma(m_1m_2+k)\over\Gamma(m_1m_2)} =
\\
= \sum_Y x^{|Y|}\left(\prod_{(i,j)\in
Y}{(m_1+j-i)(m_2+j-i)\over h(i,j)^2}\right) \equiv \sum_Y
x^{|Y|}{\cal Z}_Y^{U(1)} \ee
where for the $(i,j)\in Y=(k_1\geq k_2
\geq\ldots\geq k_l)$ with co-ordinates $(i,j)$, such that
$i=1,\ldots,l$ and $j=1,\ldots,k_i$,
the "hook" length $h(i,j)$ is
 \be \label{boxdef} h(i,j) =
k_i(Y)-j+k_j(Y^T)-i+1 \ee so that \be \prod_{(i,j)\in Y}(m_f+j-i)
= \prod_{i=1}^l\prod_{j=1}^{k_i}(m_f+j-i) =
\prod_{i=1}^l(m_f+1-i)\ldots (m_f+k_i-i) =
\\
= \prod_{i=1}^l{\Gamma\left(m_f+k_i-i\right)\over\Gamma\left(m_f+1-i\right)},
\ \ \ \ \ \ \ f=1,2
\ee
Coefficients of expansion \rf{c1expn} have already typical shape of Nekrasov's formulas,
where (in the notations of \cite{agt})
\be
{\cal Z}_Y^{U(1)} = \prod_{(i,j)\in Y}
\frac{\phi_Y(m_1;i,j)\phi_Y(m_2;i,j)}{E^2_Y(0;i,j)}
\label{u1agt}
\ee
with
\be\label{phiE}
\phi_Y(m;i,j) = m+i-j, \\
E_Y(\alpha;i,j) = \alpha+(k^T_j-i+1) +(k_i-j)
\ee
Above $\{k_i\}$ and $\{k^T_j\}$ are respectively the lengths of rows
and heights of columns in the Young diagram $Y$ and its transposed $Y^T$.

Formulas \rf{c1expn}-\rf{u1agt} follow immediately from the Cauchy formula for the Schur
functions (see e.g. \cite{McD}, for recent reviews of the
character decompositions see also \cite{chardeco1,LMN,chardeco})
\be
\label{cauchi}
\prod_{i,j}{1\over 1-\lambda_i\lambda'_j} = \sum_{Y}
s_Y(\lambda)s_Y(\lambda')
\ee
which is, when written in terms of the Miwa variables
\be
t_k = \frac{1}{k}\sum_{i=1}^N \lambda_i^k
\label{miwa}
\ee
the decomposition formula
\be
\exp\left(\sum_{k=1}^\infty
kt_kt'_k\right) = \sum_Y \chi_Y(t)\chi_Y(t'),
\label{chichi}
\ee
for the
charachers (due to the first Weyl formula)
\be
s_Y(\lambda) =
\frac{\det_{ij}\lambda_j^{k_i+N-i}}{\det_{ij} \lambda_j^{N-i}} = \chi_Y (t)
\label{Weyl2}
\ee
associated with the Young diagrams $Y$ so that $\chi_Y=\chi[k_1k_2\ldots k_l]$, e.g.
\be
\chi[0] = 1,   \\
\chi[1] = t_1,   \\
\chi[2] = t_2+\frac{t_1^2}{2}, \ \ \ \ \
\chi[11] = -t_2+\frac{t_1^2}{2},   \\
\chi[3] = t_3 + {t_1t_2} + \frac{t_1^3}{6},\ \ \ \ \
\chi[21] = -t_3  + \frac{t_1^3}{3},\ \ \ \ \
\chi[111] = t_3 - {t_1t_2} + \frac{t_1^3}{6}, \\
\ldots
\label{chars}
\ee
Literally, one has to take in \rf{cauchi} the special values
$$\{\lambda_i\} = \Big\{
\underbrace{\sqrt{x},\sqrt{x},\ldots,\sqrt{x}}_{m\ {\rm times}},
0,0,\ldots\Big\}
$$
(and analytically continue further from integer values of $m_f$, in terms of Miwa variables
\rf{miwa} this corresponds to $t_k \sim {m\over k} x^{k/2}$), then
\be
\label{schurm}
s_Y(\lambda) = x^{|Y|/2}\prod_{(i,j)\in Y}{m+j-i\over h(i,j)}
\ee
which can be treated as the case of coincident $\lambda$'s in the Weyl formula (\ref{Weyl2}),
after applying the l'H\^opital rule. Putting further $x={1\over m^2}$, and taking
the limit $m\to\infty$, one gets further
\be
\label{schupla}
d_Y \equiv \lim_{m\to\infty} m^{-|Y|}\prod_{(i,j)\in Y}{m+j-i\over h(i,j)}
=\prod_{(i,j)\in Y}{1\over h(i,j)} =
\\
= {{\rm dim}R_Y\over |Y|!} = \prod_{i<j}{k_i(Y)-k_j(Y) + j-i\over j-i}
\ee
whose square $\mu_Y=d_Y^2$ is known also as the Plancherel measure.
Summing over partitions in (\ref{c1expn}) is
performed, using particular cases of the formulas
\rf{cauchi}, \rf{chichi} (the Burnside theorems)
\be
\sum_Y d_Y\chi_Y(t) = e^{t_1},\ \ \  ({\rm at}\ t'_k=\delta_{k,1})
\\
\sum_Y \mu_Y x^{|Y|} = \sum_Y d_Y^2 x^{|Y|} = e^x, \ \ \ ({\rm if\ also}\ t_k = x\delta_{k,1})
\label{tau1}
\ee
We hereby conclude, that in the free field case, there exists a combinatorial re-interpretation
of the decomposition for conformal block
\fr{
\label{BvsZab}
\sum_{Y,Y'}x^{|Y|}{\cal B}_{Y,Y'}^{\alpha_1+\alpha_2}
(\alpha_1,\alpha_2,\alpha_3,\alpha_4)
= (1-x)^{-2\alpha_1\alpha_3} = (1-x)^{-m_1m_2}
= \sum_Y  x^{|Y|}{\cal Z}_Y^{U(1)}
}
where the coefficients ${\cal Z}_Y^{U(1)}$ are defined in \rf{u1agt}.

%Note also, that all these formulas exist in fact
%for arbitrary $\epsilon_{1,2}$, i.e. for $\epsilon\neq 0$ (cf.
%with \cite{NY}), since the nontrivial dependence upon $\epsilon_{1,2}$ at fixed dimensions
%$\Delta_i$ of external fields is absorbed into the charges $\alpha_i$ in \rf{cbQ0} (or $m_f$
%in \rf{c1expn}).

\subsection{Beyond free fields}

Arbitrary conformal model can be effectively described
in terms of free fields \cite{fref,gmmos}, however the
number of fields can be greater than one and the
constraints (\ref{ffcons}) on the intermediate state
should be released.
The natural question is what happens then
to eqs.(\ref{B1free})-(\ref{Bmfree}).
The AGT conjecture \cite{agt} is that for generic $\alpha$,
not obligatory equal to $\alpha_1+\alpha_2$,
the coefficients
\be
{\cal B}^{(k)} = \sum_{|Y|=|Y'|=k} {\cal B}_{Y,Y'}
\ee
are equal to
the expansion coefficients of Nekrasov's functions (to be explicitly defined in s.4.2):
\fr{
\sum_{|Y|=|Y'|} x^{|Y|} {\cal B}_{Y,Y'}^\alpha(\alpha_1,\alpha_2;\alpha_3,\alpha_4)
= \sum_{Y,Y'}x^{|Y|}{\cal B}_{Y,Y'}^{\alpha_1+\alpha_2}(\alpha_1,\alpha_2;\alpha_3,\alpha_4)
\sum_{Y,Y'} x^{|Y|+|Y'|} {\cal Z}^{SU(2)}_{Y,Y'} =
\\
=
(1-x)^{-\nu} \sum_{Y,Y'} x^{|Y|+|Y'|} {\cal Z}^{SU(2)}_{Y,Y'}
= \sum_{Y,Y'} x^{|Y|+|Y'|} {\cal Z}^{U(2)}_{Y,Y'}
\label{BvsZ}
}
where the $U(1)$ factor $(1-x)^{-\nu}$ is itself represented
by the similar pattern formula (\ref{BvsZab}).
Among other things the first line of (\ref{BvsZ}) implies that
the combination of $SU(2)$ ${\cal Z}$-functions is strictly unity
whenever $\alpha = \alpha_1+\alpha_2$.
Restriction to $SU(2)$ at the r.h.s. of (\ref{BvsZ}) looks to be
related to restriction to a single free field at the l.h.s.
Note that (\ref{BvsZ}) is a universal group-theory relation
with no reference to particular conformal model at the l.h.s.
and particular SUSY Yang-Mills theory at the r.h.s.:
it relates explicit group-theoretical quantities, canonically
associated with the Young diagrams.
The only point is that though canonical,
these associations still look rather
sophisticated and lack clear interpretation
in representation theory of linear and symmetric groups.

\section{AGT relations}

The claim of \cite{agt} is actually more general than (\ref{BvsZ}).
The observation is that Nekrasov partition functions factorize exactly in the same way
as conformal blocks in (\ref{cobfac}):
\be
Z_{Nek}= Z_{cl}Z_{pert}Z_{inst}
\ee
where
\be
Z_{inst} = {\cal B}_\Delta(\Delta_1,\Delta_2;\Delta_3,\Delta_4|x)
\label{instB}
\ee
for some choice of the dimensions
$\Delta_1,\Delta_2,\Delta_3,\Delta_4$ and $\Delta$,
\be
Z_{cl} = x^\sigma
\label{class}
\ee
and
\be
Z_{pert} =
C_{\alpha_1\alpha_2}^a{\cal K}_aC_{\alpha_3\alpha_4}^a
\label{pertC}
\ee
for some choice of conformal theory and its primary states.

More than that,
this conformal theory has been claimed to be the Liouville model
with the central charge
\be
c = 1+ 6Q^2 =
1 + \frac{6(\epsilon_1+\epsilon_2)^2}{\epsilon_1\epsilon_2}
\label{ce}
\ee
and dimensions\footnote{
Note rescaling by $\epsilon_1\epsilon_2$
as compared to the standard notation (\ref{ffdims}),
it is innocent, but simplifies the formulas.
If $Q$ is parameterized in the CFT-standard way as $Q=b+1/b$
then $b = \sqrt{\epsilon_1/\epsilon_2}$.
We emphasize that we do {\it not} see any need to require that
$\epsilon_1\epsilon_2 = 1$ for identification of conformal
blocks and Nekrasov's partition functions for instanton sums.}
\be
\Delta_\alpha = \frac{\alpha(\epsilon - \alpha)}{\epsilon_1\epsilon_2}
\label{dime}
\ee
with $\epsilon = \epsilon_1+\epsilon_2$.
The $\alpha$'s are linear
combinations of parameters -- gauge fields $a$ and masses $\mu$ --
of Nekrasov's partition functions, see below.

Furthermore, this identification has been generalized to higher-point conformal
blocks: to be related to Nekrasov's functions for non-trivial
patterns of branes (quiver gauge theories).
The relations of \cite{agt} involves only Nekrasov's functions
for $SU(2)$ and $U(2)$ "gauge" groups.
As already mentioned,
for arbitrary groups it should presumably involve
conformal blocks of the reduced WZNW model \cite{gmmos}
(Affine Toda, to be concrete, \cite{agtplus}),
expressed through more than a single scalar field.

Note that while (\ref{instB}) is just the universal
relation (\ref{BvsZ}), eq.(\ref{pertC}) essentially involves
sophisticated expressions for the structure constants in the Liouville
theory
(the DOZZ vertices \cite{DOZZ}), which look like non-trivial
quantities from representation theory of the quantum Kac-Moody
algebras (see also \cite{Ligr}).
We leave (\ref{class}) and (\ref{pertC}) beyond the scope
of this paper: instead, we are going to concentrate on (\ref{instB}) in
the form of (\ref{BvsZ}).

\section{Nekrasov partition functions}

These are the
newly discovered special functions ${\cal Z}_{Y,Y'}[\epsilon_1,\epsilon_2]$
with increasing number of applications in modern mathematical
physics. For $\epsilon_1=-\epsilon_2\equiv\hbar$ they are closely related
to characters of symmetric and linear groups, if further
$\hbar\rightarrow 0$ they reproduce the Seiberg-Witten prepotentials.
The conjecture of \cite{agt} can be considered as one more application:
Nekrasov instanton partition functions describe the conformal
blocks with Virasoro in the role of the chiral algebra.
The number of $\epsilon$-parameters can be increased in this way, and these
extra parameters can actually play a role
in generalizations to other chiral algebras.
Representation-theory interpretation of generic Nekrasov
functions remains obscure, and they can be (temporarily?) considered as providing
{\it combinatorial} rather than {\it character} decompositions.

\subsection{$U(1)$ case, ${\cal Z}_{Y}$}

For particular case $\epsilon=0$, for the $c=1$ free conformal theory,
eq.(\ref{BvsZab}) is the well known $U(1)$ precursor of the AGT
relation (\ref{BvsZ}).
It describes Nekrasov partition function of the $U(1)$ gauge theory
with $N_f=2$ flavours as
a conformal block \cite{LMN}
\be
\label{u1cb}
{\cal Z}_{inst}^{U(1),N_f=2}(x;m_1,m_2) =
\langle e^{i (\alpha + m_2)\varphi(\infty) } e^{-i m_2 \varphi (1) }
x^{L_0} e^{i m_1 \varphi(1)}e^{ - i (\alpha+m_1)\varphi(0)} \rangle =
\\
= x^{m_1^2+(\alpha+m_1)^2}\langle e^{i (\alpha + m_2)\varphi(\infty) } e^{-i m_2 \varphi (1) }
e^{i m_1 \varphi(x)}e^{ - i (\alpha+m_1)\varphi(0)} \rangle =
x^{\alpha^2}\left(1-x\right)^{-m_1m_2}
\ee
where for simplicity $\epsilon_1=-\epsilon_2=\hbar=1$.
As expected, the intermediate channel for the conformal block is projected
to a single representation with dimension of the primary field
$\Delta_\alpha \sim \alpha^2$. Conformal block \rf{u1cb} has an expansion
\rf{c1expn} over a single set of Young diagramms, we already discussed in
sect.~\ref{ss:comb}.

Formula \rf{u1cb} is a
matrix element in the $\widehat{U(1)}$ conformal theory with $c=1$,
which possesses not only a free-boson but also a
free-fermion representations \cite{LMN,NO} for the current
\be
\label{u1c}
J(z) = \partial\varphi(z) = :\tilde\psi(z)\psi(z): = \sum_{n\in\mathbb{Z}}{J_n\over z^{n+1}}
\ee
so that the stress-energy tensor (\ref{ste})
is just $T(z) \sim J(z)^2$.

The limit of infinite masses $m_{1,2}\to\infty$ in \rf{u1cb}
corresponds to decoupling of matter from the gauge fields
and leads to
\be
\label{zu1}
Z^{U(1)}(\alpha,\Lambda) = \langle \alpha|\ e^{J_1} \Lambda^{L_0} e^{J_{-1}}|\alpha\rangle =
\Lambda^{\alpha^2/2}\sum_Y \mu_Y\Lambda^{|Y|}\ \stackrel{\rf{tau1}}{=}\
\Lambda^{\alpha^2/2}e^{\Lambda}
\ee
with $\Lambda^2 = m_1m_2x^2={\rm fixed}$ at $x\to 0$, $m_{1,2}\to\infty$ being the scale
parameter of the pure $U(1)$ gauge theory.

As was already noticed in sect.~\ref{ss:comb}, these formulas remain intact
at arbitrary $\epsilon\neq 0$, since switching on $\epsilon$ corresponds
to nothing more than the ``twisting'' of the stress-tensor of the
$\widehat{U(1)}$ theory $T(z)\to T(z)+\epsilon\d J(z)$. This just shifts
the weights of the twisted fermions and deforms the combinatorial formulas,
whose summation leads basically to the same results (see e.g. \cite{NY}).

\subsection{$SU(2)$ case, ${\cal Z}_{Y,Y'}$}

The $SU(2)$ Nekrasov's functions ${\cal Z}_{Y,Y'}$ are much more involved.
They are manifestly given by the formulas
\be
{\cal Z}_{Y,Y'}={\eta(Y_1,Y_2)\over \xi(Y_1,Y_2)}
\ee
with
\be\label{etaU}
\eta(Y_1,Y_2)=\prod_{(i,j)\in Y_1}\prod_{\alpha=1}^4
\Big(\phi(a_1,i,j)+\mu_\alpha\Big)
\prod_{(i,j)\in Y_2}\prod_{\alpha=1}^4
\Big(\phi(a_2,i,j)+\mu_\alpha\Big)
\ee
\be
\xi(Y_1,Y_2)=\prod_{(i,j)\in Y_1}E(a_1-a_1,Y_1,Y_1,i,j)
\Big(\epsilon-E(a_1-a_1,Y_1,Y_1,i,j)\Big)\times\\
\times\prod_{(i,j)\in Y_1}E(a_1-a_2,Y_1,Y_2,i,j)
\Big(\epsilon-E(a_1-a_2,Y_1,Y_2,i,j)\Big)\times\\
\times\prod_{(i,j)\in Y_2}E(a_2-a_1,Y_2,Y_1,i,j)
\Big(\epsilon-E(a_2-a_1,Y_2,Y_1,i,j)\Big)\times\\
\times\prod_{(i,j)\in Y_2}E(a_2-a_2,Y_2,Y_2,i,j)
\Big(\epsilon-E(a_2-a_2,Y_2,Y_2,i,j)\Big)
\ee
where $a_2=-a_1$ and
\be
\phi(a,i,j)=a+\epsilon_1(i-1)+\epsilon_2(j-1)\\
E(a,Y_1,Y_2,i,j)=a+\epsilon_1\Big(k^T_j(Y_1)-i+1\Big)-\epsilon_2\Big(k_i(Y_2)-j\Big)
\ee
defined for two Young diagrams $Y_1$ and $Y_2$.
These functions are natural generalization of the
$U(1)$ quantities in (\ref{phiE}). Note that for $U(1)$ $\epsilon_1=
-\epsilon_2$ and the scaling factor $\epsilon_1\epsilon_2$ has not been
introduced yet in (\ref{phiE}). Note also that, in variance with the
$U(1)$ case, where $\sqrt{2}\alpha=m$ enters the product in the numerator of
(\ref{phiE}), in the $SU(2)$ case, (\ref{etaU}) there are no square
roots in front of $a_i$.

Explicit expressions for the first few terms of their expansion are
\be
{\cal Z}_{[1][0]} = -\frac{1}{\epsilon_1\epsilon_2}\cdot
\frac{\prod_{r=1}^4 (a + \mu_r)}
{2a(2a+\epsilon)},\\
{\cal Z}_{[0][1]} = -\frac{1}{\epsilon_1\epsilon_2}\cdot
\frac{\prod_{r=1}^4 (a - \mu_r)}
{2a(2a-\epsilon)};
\label{Z1}
\ee
for the instantonic charge $k=|Y|+|Y'|=1$,
\be
{\cal Z}_{[2][0]} = \frac{1}{2!\,\epsilon_1\epsilon_2^2
(\epsilon_1-\epsilon_2)}\cdot
\frac{\prod_{r=1}^4 (a + \mu_r)(a+\mu_r+\epsilon_2)}
{2a(2a+\epsilon_2)(2a+\epsilon)(2a+\epsilon+\epsilon_2)},\\
{\cal Z}_{[0][2]} = \frac{1}{2!\,\epsilon_1\epsilon_2^2
(\epsilon_1-\epsilon_2)}\cdot
\frac{\prod_{r=1}^4 (a - \mu_r)(a-\mu_r-\epsilon_2)}
{2a(2a-\epsilon_2)(2a-\epsilon)(2a-\epsilon-\epsilon_2)},\\
{\cal Z}_{[11][0]} = -\frac{1}{2!\,\epsilon_1^2\epsilon_2
(\epsilon_1-\epsilon_2)}\cdot
\frac{\prod_{r=1}^4 (a + \mu_r)(a+\mu_r+\epsilon_1)}
{2a(2a+\epsilon_1)(2a+\epsilon)(2a+\epsilon+\epsilon_1)},  \\
{\cal Z}_{[0][11]} = -\frac{1}{2!\,\epsilon_1^2\epsilon_2
(\epsilon_1-\epsilon_2)}\cdot
\frac{\prod_{r=1}^4 (a - \mu_r)(a-\mu_r-\epsilon_1)}
{2a(2a-\epsilon_1)(2a-\epsilon)(2a-\epsilon-\epsilon_1)}, \\
{\cal Z}_{[1][1]} = \frac{1}{\epsilon_1^2\epsilon_2^2}\cdot
\frac{\prod_{r=1}^4 (a + \mu_r)(a-\mu_r)}
{(4a^2-\epsilon_1^2)(4a^2-\epsilon_2^2)};
\label{Z2}
\ee
for $k=2$, then
\be
{\cal Z}_{[3][0]} = -\frac{1}{3!\,\epsilon_1\epsilon_2^3
(\epsilon_1-\epsilon_2)(\epsilon_1-2\epsilon_2)}\cdot
\frac{\prod_{r=1}^4 (a + \mu_r)(a+\mu_r+\epsilon_2)(a+\mu_r+2\epsilon_2)}
{2a(2a+\epsilon_2)(2a+2\epsilon_2)(2a+\epsilon)(2a+\epsilon+\epsilon_2)
(2a+\epsilon+2\epsilon_2)},\\
etc.
\label{Z3}
\ee
Clearly, there is a symmetry
\be
{\cal Z}_{Y',Y}(a,\epsilon_1,\epsilon_2) =
{\cal Z}_{Y,Y'}(-a,\epsilon_1,\epsilon_2)
\ee
and
\be
{\cal Z}_{Y,Y'}(a,\epsilon_1,\epsilon_2)
={\cal Z}_{Y^T,{Y'}^T}(a,\epsilon_2,\epsilon_1)
\ee
The mnemonic rule to construct these expressions is rather simple:
in transition $[n]\rightarrow [n+1]$ one introduces additional
entries with an extra $\epsilon_2$ both in the numerator and denominator,
while in transition $[1^n]\rightarrow [1^{n+1}]$ one adds extra
$\epsilon_1$.

\section{Beyond free fields, continued}

\subsection{The case of
$\alpha_1=\alpha_2=\alpha_3=\alpha_4=0$: the conformal block}

We now return to consideration of conformal blocks ${\cal B}_\Delta$
with unconstrained dimension $\Delta_\alpha$.
The simplest case to begin with is when all the other
dimensions are vanishing, i.e. the four "external" primaries
are just unit operators with
$\Delta_1 = \Delta_2=\Delta_3=\Delta_4=0$, while the "intermediate"
$\Delta = \frac{\alpha(\epsilon-\alpha)}{\epsilon_1\epsilon_2}$ remains nontrivial
and arbitrary.
Then, formulas (\ref{B1})-(\ref{B3}) reduce to:
\be
{\cal B}^{(1)}(0,0;0,0) \ \stackrel{(\ref{B1})}{=}\  \frac{\Delta}{2}
\ \stackrel{(\ref{dime})}{=}\
\frac{\alpha(\epsilon-\alpha)}{2\epsilon_1\epsilon_2}
= -\frac{4a^2-\epsilon^2}{8\epsilon_1\epsilon_2},
\label{B10}
\ee
\be
{\cal B}^{(2)}(0,0;0,0) \ \stackrel{(\ref{B2})}{=}\
\frac{\Delta\Big(8\Delta^3+(c+8)\Delta^2+(2c-8)\Delta+c\Big)}
{4\Big(16\Delta^2+(2c-10)\Delta+c\Big)} =   \\
= \frac{(4a^2-\epsilon^2)
}{256
\epsilon_1^2\epsilon_2^2\Big(4a^2-(2\epsilon_1+\epsilon_2)^2\Big)
\Big(4a^2-(\epsilon_1+2\epsilon_2)^2\Big)}\cdot \\
\Big(128 a^6
-48a^4(4\epsilon_1^2 + 11\epsilon_1 \epsilon_2 + 4\epsilon_2^2)
+24 a^2(3\epsilon_1^4 + 23 \epsilon_1^3 \epsilon_2
+ 36\epsilon_1^2\epsilon_2^2
+23 \epsilon_1 \epsilon_2^3  + 3 \epsilon_2^4) -  \\
-(8 \epsilon_1^6+105 \epsilon_1^5 \epsilon_2+420 \epsilon_1^4 \epsilon_2^2
+662 \epsilon_1^3 \epsilon_2^3+420 \epsilon_1^2 \epsilon_2^4
+105 \epsilon_1 \epsilon_2^5+8 \epsilon_2^6)\Big),
\label{B20}
\ee
\be
{\cal B}^{(3)}(0,0;0,0) \ \stackrel{(\ref{B3})}{=}\
\frac{\Delta(\Delta+2)\Big(8\Delta^3+(c+18)\Delta^2 +
(3c-14)\Delta + 2c\Big)}{24\Big(16\Delta^2+(2c-10)\Delta+c\Big)}= \\
= -\frac{(4a^2-\epsilon^2)
\Big(4a^2-(\epsilon_1^2+10\epsilon_1\epsilon_2+\epsilon_2^2)\Big)
}
{6144\epsilon_1^3\epsilon_2^3\Big(4a^2-(2\epsilon_1+\epsilon_2)^2\Big)
\Big(4a^2-(2\epsilon_1+\epsilon_2)^2\Big)}\cdot \\
\cdot
\Big(128 a^6
-16a^4(12 \epsilon_1^2 +43 \epsilon_1 \epsilon_2 +12 \epsilon_2^2 )
+ 8a^2(9\epsilon_1^4 + 91\epsilon_1^3 \epsilon_2
+142 \epsilon_1^2 \epsilon_2^2 + 91\epsilon_1 \epsilon_2^3
+ 9\epsilon_2^4) -   \\
-(8 \epsilon_1^6+139 \epsilon_1^5 \epsilon_2
+632 \epsilon_1^4 \epsilon_2^2+1034 \epsilon_1^3 \epsilon_2^3
+632 \epsilon_1^2 \epsilon_2^4 +139 \epsilon_1 \epsilon_2^5
+8 \epsilon_2^6)\Big)
\label{B30}
\ee
where at the last step we made a shift
\be
\alpha = a+\frac{\epsilon}{2},
\ee
which symmetrizes and slightly simplifies the formulas.
Obviously, denominators (Kac determinants) in (\ref{B1})-(\ref{B3})
are nicely consistent with the substitutions (\ref{ce})
and (\ref{dime}) -- a fact, very well known from generic
conformal field theory.
Note also that denominators are independent of the
other dimensions $\Delta_1,\ldots,\Delta_4$, thus this
consistency will persist for generic 4-point conformal blocks.

However, the numerators in (\ref{B10})-(\ref{B30})
look pretty sophisticated.
The question is: what is the appropriate analogue of character
decomposition (\ref{BvsZab}) for these quantities?
The AGT answer is (\ref{BvsZ}) and we now proceed to the check and
analysis of this relation.

\subsection{The case of $\alpha_1=\alpha_2=\alpha_3=\alpha_4=0$: Nekrasov's functions}

In this case we put $\nu = 0$ in (\ref{BvsZ}).
Now we need to adjust the four parameters $\mu_1,\ldots,\mu_4$
so that (\ref{BvsZ}) is satisfied for explicit expressions
(\ref{B10})-(\ref{B30}) and (\ref{Z1})-(\ref{Z3}).

First of all, we need to match (\ref{B10}) and (\ref{Z1}):
\be
{\cal B}^{(1)}(0,0;0,0)
=-\frac{4a^2-\epsilon^2}{8\epsilon_1\epsilon_2} =
{\cal Z}_{[1][0]} + {\cal Z}_{[0][1]} =
-\frac{1}{\epsilon_1\epsilon_2}\cdot
\frac{\prod_{r=1}^4 (a + \mu_r)}{2a(2a+\epsilon)}
-\frac{1}{\epsilon_1\epsilon_2}\cdot
\frac{\prod_{r=1}^4 (a - \mu_r)}{2a(2a-\epsilon)}
\label{BZ1}
\ee
In order to get rid of poles at the r.h.s., one of the $\mu$
parameters should be equal to $\epsilon/2$, let it be $\mu_4=\epsilon/2$.
Then the r.h.s. of (\ref{BZ1}) turns into
\be
-\frac{a^2+\mu_1\mu_2+\mu_1\mu_3+\mu_2\mu_3}{2\epsilon_1\epsilon_2}
\ee
and the matching condition with the l.h.s. is
\be
s_2\equiv\mu_1\mu_2+\mu_1\mu_3+\mu_2\mu_3 = -\frac{\epsilon^2}{4}
\ee

Far less trivial is the matching between (\ref{B20}) and (\ref{Z2}):
\be
{\cal B}^{(2)}(0,0;0,0)=
\frac{(4a^2-\epsilon^2)
}{256
\epsilon_1^2\epsilon_2^2\Big(4a^2-(2\epsilon_1+\epsilon_2)^2\Big)
\Big(4a^2-(2\epsilon_1+\epsilon_2)^2\Big)}
\Big(128 a^6
-48a^4(4\epsilon_1^2 + 11\epsilon_1 \epsilon_2 + 4\epsilon_2^2) + \\
+24 a^2(3\epsilon_1^4 + 23 \epsilon_1^3 \epsilon_2
+ 36\epsilon_1^2\epsilon_2^2
+23 \epsilon_1 \epsilon_2^3  + 3 \epsilon_2^4)-\\
-(8 \epsilon_1^6+105 \epsilon_1^5 \epsilon_2+420 \epsilon_1^4 \epsilon_2^2
+662 \epsilon_1^3 \epsilon_2^3+420 \epsilon_1^2 \epsilon_2^4
+105 \epsilon_1 \epsilon_2^5+8 \epsilon_2^6)\Big) =   \\
= {\cal Z}_{[2][0]} + {\cal Z}_{[0][2]} +
{\cal Z}_{[11][0]} + {\cal Z}_{[0][11]} + {\cal Z}_{[1][1]} =
\frac{1}{\epsilon_1^2\epsilon_2^2}\cdot
\frac{\prod_{r=1}^4 (a + \mu_r)(a-\mu_r)}
{(4a^2-\epsilon_1^2)(4a^2-\epsilon_2^2)} +   \\
+\frac{1}{2!\,\epsilon_1\epsilon_2^2
(\epsilon_1-\epsilon_2)}\cdot
\frac{\prod_{r=1}^4 (a + \mu_r)(a+\mu_r+\epsilon_2)}
{2a(2a+\epsilon_2)(2a+\epsilon)(2a+\epsilon+\epsilon_2)}
+ \frac{1}{2!\,\epsilon_1\epsilon_2^2
(\epsilon_1-\epsilon_2)}\cdot
\frac{\prod_{r=1}^4 (a - \mu_r)(a-\mu_r-\epsilon_2)}
{2a(2a-\epsilon_2)(2a-\epsilon)(2a-\epsilon-\epsilon_2)} - \\
-\frac{1}{2!\,\epsilon_1^2\epsilon_2
(\epsilon_1-\epsilon_2)}\cdot
\frac{\prod_{r=1}^4 (a + \mu_r)(a+\mu_r+\epsilon_1)}
{2a(2a+\epsilon_1)(2a+\epsilon)(2a+\epsilon+\epsilon_1)}
-\frac{1}{2!\,\epsilon_1^2\epsilon_2
(\epsilon_1-\epsilon_2)}\cdot
\frac{\prod_{r=1}^4 (a - \mu_r)(a-\mu_r-\epsilon_1)}
{2a(2a-\epsilon_1)(2a-\epsilon)(2a-\epsilon-\epsilon_1)}  =   \\
= \frac{128a^{14} + \ldots}
{\epsilon_1^2\epsilon_2^2(4a^2-\epsilon_1^2)(4a^2-\epsilon_2^2)
(4a^2-\epsilon^2)\Big(4a^2-(2\epsilon_1+\epsilon_2)^2\Big)
\Big(4a^2-(\epsilon_1+2\epsilon_2)^2\Big)}
\ee
Numerator at the r.h.s. is a polynomial of degree $7$ in $a^2$
and it should match the l.h.s., which we multiply by
$(4a^2-\epsilon^2)(4a^2-\epsilon_1^2)(4a^2-\epsilon_2^2)$
in order to convert it into a similar polynomial of degree $14$
with the first term $128a^{14}$.
The next coefficients of these two polynomials are:
$$
\begin{array}{c|c|c}
{\rm degree} & {\rm l.h.s.} & {\rm r.h.s.} \\
\hline
a^{14}: & 128 & 128 \\
\hline
a^{12}: & -288\epsilon^2 - 80\epsilon_1\epsilon_2
& 256\sum_{r<s}^4 \mu_r\mu_s - 128\epsilon \sum_{r=1}^4 \mu_r
- 160\epsilon^2-80\epsilon_1\epsilon_2 \\
&& = 256 \sum_{r<s}^3 \mu_r\mu_s
-224\epsilon^2 -80\epsilon_1\epsilon_2\\
\hline
&&\ldots
\end{array}
$$
Further lines in the right column are quite involved. For instance,
the coefficient in front of $a^{10}$ is
\be
16\times\left[6s_4+8s_2^2-8\epsilon
(s_1s_2+s_3)-(16\epsilon_1^2+16\epsilon_2^2 +38\epsilon_1\epsilon_2
)s_2+(2\epsilon_1^2+2\epsilon_2^2+
3\epsilon_1\epsilon_2)s_1^2+\right.\\\left.
+(8\epsilon_1^3+8\epsilon_2^3+ 29\epsilon\epsilon_1\epsilon_2
)s_1+(30\epsilon_1\epsilon_2+
18\epsilon_1^2+18\epsilon_2^2)\epsilon_1\epsilon_2 \right]
\ee
where
$s_k$ denote the symmetric polynomials of four $\mu_i$ of degree
$k$, $s_4\equiv \mu_1\mu_2\mu_3\mu_4$,
$s_3=\sum_{a>b>c}\mu_a\mu_b\mu_c$ etc.
We do not write down its left-column counterpart and the further
lines here: expressions are getting pretty long, while the procedure
is hopefully already clear.

The last line in the right column is obtained after substitution of
$\mu_4=\epsilon/2$. It coincides with the left column when choosing
the following $\mu$'s: $\mu_{1,2}=\pm\epsilon/2$, $\mu_{3,4} =
\mp\epsilon/2$, or $\mu_{1,2}=-\epsilon/2$, $\mu_{3,4} =
\epsilon/2$, or $\mu_{1}=-\epsilon/2$, $\mu_{2} = 3\epsilon/2$, $\mu_{3,4} =
\epsilon/2$ (definitely, with all permutations of $\mu$'s). The
reason for existence of several solutions will become clear in the
next subsection. With this choice the coefficients of all other
powers of $a^2$ also match in the two columns.

Analogous check works also at level $3$.

\subsection{Restoring dimensions $\Delta_1,\ldots,\Delta_4$}

It is now a simple exercise to switch on non-vanishing
$\alpha_1,\ldots,\alpha_4$. Start with the first level:
\be
{\cal
B}^{(1)}(\Delta_1,\Delta_2;\Delta_3,\Delta_4)=-
{2a^4+2a^2\Big(\epsilon(-\alpha_1+\alpha_2-\alpha_3+\alpha_4)+
\alpha_1^2-\alpha_2^2+\alpha_3^2-\alpha_4^2
-\epsilon^2/2\Big)\over \epsilon_1\epsilon_2(4a^2-\epsilon^2)}-\\
-{2\Big(\alpha_1(\epsilon-\alpha_1)-\alpha_2(\epsilon-\alpha_2)+\epsilon^2/4\Big)
\Big(\alpha_3(\epsilon-\alpha_3)-\alpha_4(\epsilon-\alpha_4)+\epsilon^2/4\Big)
\over \epsilon_1\epsilon_2(4a^2-\epsilon^2)}
\ee
to be compared with
\be
{\cal Z}_{[1][0]} + {\cal Z}_{[0][1]}+\nu=-
{2a^4+(2s_2-\epsilon s_1-4\epsilon_1\epsilon_2\nu)a^2+2s_4-\epsilon
s_3 +\epsilon_1\epsilon_2\epsilon^2\nu\over
\epsilon_1\epsilon_2(4a^2-\epsilon^2)}
\ee
This gives two relations
to determine four $\mu$'s and $\nu$:
\be
2\Big(\epsilon(-\alpha_1+\alpha_2-\alpha_3+\alpha_4)+
\alpha_1^2-\alpha_2^2+\alpha_3^2-\alpha_4^2 -\epsilon^2/2\Big)=
2s_2-\epsilon s_1-4\epsilon_1\epsilon_2\nu\\
2\Big(\alpha_1(\epsilon-\alpha_1)-\alpha_2(\epsilon-\alpha_2)+\epsilon^2/4\Big)
\Big(\alpha_3(\epsilon-\alpha_3)-\alpha_4(\epsilon-\alpha_4)+\epsilon^2/4\Big)=
2s_4-\epsilon s_3 +\epsilon^2\epsilon_1\epsilon_2\nu
\ee
Three more
relations are obtained from level two - those being more
involved, and also at level three.

However, despite the number of emerging matching constraints is large,
they are all consistent(!), this is exactly the observation of \cite{agt},
and they do have a common solution.
In fact, there are {\em eight} common solutions
(of course, modulo $24$ permutations of $\mu_1,\ldots,\mu_4$):
\fr{
I:\ \
\mu_1=-{\epsilon\over 2}+\alpha_1+\alpha_2, \ \ \mu_2={\epsilon\over
2}+\alpha_1-\alpha_2, \ \ \mu_3=-{\epsilon\over
2}+\alpha_3+\alpha_4, \ \ \
\mu_4={\epsilon\over 2}+\alpha_3-\alpha_4,
 \\
\nu = \frac{2\alpha_1\alpha_3}{\epsilon_1\epsilon_2}\nonumber
}
\be\label{solutions}
II:\ \ \mu_1={3\epsilon\over 2}-\alpha_1-\alpha_2, \ \
\mu_2={\epsilon\over 2}+\alpha_2-\alpha_1, \ \ \mu_3={3\epsilon\over
2}-\alpha_3-\alpha_4, \ \ \
\mu_4={\epsilon\over 2}+\alpha_4-\alpha_3, \\
\nu =
\frac{2(\epsilon-\alpha_1)(\epsilon-\alpha_3)}{\epsilon_1\epsilon_2}
\\
III:\ \ \mu_1=-{\epsilon\over 2}+\alpha_1+\alpha_2, \ \
\mu_2={\epsilon\over 2}+\alpha_1-\alpha_2, \ \ \mu_3={3\epsilon\over
2}-\alpha_3-\alpha_4, \ \ \
\mu_4={\epsilon\over 2}+\alpha_4-\alpha_3, \\
\nu = \frac{2\alpha_1(\epsilon-\alpha_3)}{\epsilon_1\epsilon_2}
\\
IV:\ \ \mu_1={\epsilon\over 2}+\alpha_2-\alpha_1, \ \
\mu_2={3\epsilon\over 2}-\alpha_1-\alpha_2, \ \ \mu_3=-{\epsilon\over
2}+\alpha_3+\alpha_4, \ \ \
\mu_4={\epsilon\over 2}+\alpha_3-\alpha_4, \\
\nu = \frac{2(\epsilon-\alpha_1)\alpha_3}{\epsilon_1\epsilon_2}
\ee
\be\label{solutions1}
V:\ \ \mu_1=-{\epsilon\over 2}+\alpha_1+\alpha_2, \ \
\mu_2={\epsilon\over 2}+\alpha_2-\alpha_1, \ \ \mu_3=-{\epsilon\over
2}+\alpha_3+\alpha_4, \ \ \
\mu_4={\epsilon\over 2}+\alpha_4-\alpha_3, \\
\nu = \frac{(\alpha_2+\alpha_4)^2-\alpha_1^2-\alpha_3^2-(\alpha_2
+\alpha_4-\alpha_1-\alpha_3)\epsilon}{\epsilon_1\epsilon_2}
\\
VI:\ \ \mu_1={3\epsilon\over 2}-\alpha_1-\alpha_2, \ \
\mu_2={\epsilon\over 2}+\alpha_1-\alpha_2, \ \ \mu_3=-{\epsilon\over
2}+\alpha_3+\alpha_4, \ \ \
\mu_4={\epsilon\over 2}+\alpha_4-\alpha_3, \\
\nu =  \frac{(\alpha_2-\alpha_4)^2-\alpha_1^2-\alpha_3^2+(\alpha_4
-\alpha_2+\alpha_1+\alpha_3)\epsilon}{\epsilon_1\epsilon_2}
\\
VII:\ \ \mu_1=-{\epsilon\over 2}+\alpha_1+\alpha_2, \ \
\mu_2={\epsilon\over 2}+\alpha_2-\alpha_1, \ \ \mu_3={3\epsilon\over
2}-\alpha_3-\alpha_4, \ \ \
\mu_4={\epsilon\over 2}+\alpha_3-\alpha_4, \\
\nu = \frac{(\alpha_2-\alpha_4)^2-\alpha_1^2-\alpha_3^2+(\alpha_2
-\alpha_4+\alpha_1+\alpha_3)\epsilon}{\epsilon_1\epsilon_2}
\\
VIII:\ \ \mu_1={3\epsilon\over 2}-\alpha_1-\alpha_2, \ \
\mu_2={\epsilon\over 2}+\alpha_1-\alpha_2, \ \ \mu_3={3\epsilon\over
2}-\alpha_3-\alpha_4, \ \ \
\mu_4={\epsilon\over 2}+\alpha_3-\alpha_4, \\
\nu =
\frac{2\epsilon^2+(\alpha_2+\alpha_4)^2-\alpha_1^2-\alpha_3^2+(\alpha_1
+\alpha_3-3\alpha_2-3\alpha_4)\epsilon}{\epsilon_1\epsilon_2} \ee
Given these values, it is easy to check consistency of (\ref{BvsZ}) at levels
two and three. One of these solutions (III)
coincides with that in
\cite{agt,agtplus} (after appropriately shifting $\mu$'s and changing notations:
$\alpha_1\leftrightarrow\alpha_2$).
Equality in the first line of (\ref{BvsZ}) corresponds to choice
of the boxed solution in (\ref{solutions}), for other choices one should
substitute $\alpha_1+\alpha_2$ appropriately.

Hence, it is indeed quite easy to believe that this correspondence survives at all
higher levels, but the check requires either a  tedious calculation or a clever theoretical proof
- which should be the natural next step in the study of the
AGT conjecture.

\subsection{Symmetries, zeroes and poles}

Now let us discuss what is the reason for existence of eight solutions,
(\ref{solutions})-(\ref{solutions1})
for the correspondence \rf{BvsZ}. The first four solutions (\ref{solutions}) are
related to each other
by transformations of the type $\alpha\to\epsilon-\alpha$.
Indeed, $I$ turns to $II$
under all $\alpha_i\to\epsilon-\alpha_i$, $II$ goes to
$III$ under $\alpha_{1,2}
\to\epsilon-\alpha_{1,2}$ and to $IV$ under $\alpha_{3,4}\to\epsilon-
\alpha_{3,4}$. Similarly the second four solutions are related:
$V\to VI\to VII\to VIII$ is provided by the chain of transformations
$\alpha_{1,2}\to\epsilon-\alpha_{1,2}$, $\alpha_{1,2,3,4}\to\epsilon
-\alpha_{1,2,3,4}$, $\alpha_{1,2}\to\epsilon-\alpha_{1,2}$.

Other possible reflections $\alpha\to\epsilon-\alpha$ do not lead to new solutions.
Indeed, say, changing just one $\alpha_1\to\epsilon-\alpha_1$ transforms $I$ to
$IV$ etc.

Therefore, the AGT relation is invariant under reflecting
any of $\alpha_i$. This symmetry is evident at the conformal side of
the AGT relation. Indeed, the
dimensions of operators (\ref{ffdims}) are invariant under reflecting any of
$\alpha$.

More intriguing is the second, $\mathbb{Z}_2$
symmetry, which relates the first four, (\ref{solutions}) and the second
four, (\ref{solutions1}) solutions. It is generated by the permutation
$\alpha_1\leftrightarrow \alpha_2$,
$\alpha_3\leftrightarrow\alpha_4$. Note that under this transformation one should also
change the "$U(1)$ parameter" $\nu$:
\be
\frac{2\alpha_1\alpha_3}{\epsilon_1\epsilon_2}\ \to\
\frac{(\alpha_2+\alpha_4)^2-\alpha_1^2-\alpha_3^2-(\alpha_2
+\alpha_4-\alpha_1-\alpha_3)\epsilon}{\epsilon_1\epsilon_2}
\label{atr}
\ee
This transformation of $\nu$ is trivial only for the case of free fields,
i.e. when conditions (\ref{ffcons}) are imposed on $\alpha$'s:
then the r.h.s. in (\ref{atr}) coincides with the l.h.s.
Generally, this looks like a non-trivial symmetry (duality)
of the Nekrasov partition functions for conformal theories,
\be
\sum_{k=0}^m \frac{\Gamma(m-k+\nu)}{k!\Gamma(\nu)}
\sum_{|Y|+|Y'|=k} Z_{Y,Y'}\{\mu\} =
\sum_{k=0}^m \frac{\Gamma(m-k+\tilde\nu)}{k!\Gamma(\tilde\nu)}
\sum_{|Y|+|Y'|=k} Z_{Y,Y'}\{\tilde\mu\}
\ee

%\subsection{Zeroes and poles}

As we already mentioned, the first line of (\ref{BvsZ}) implies that
Nekrasov partition functions ${\cal Z}_{Y,Y'}$ with nontrivial
$Y,Y'\neq \emptyset$ should vanish
when we put $\alpha=\alpha_1+\alpha_2$ and
$\alpha_1+\ldots +\alpha_4 = \epsilon$. This can happen if
$a+\mu_i=0$ and $-a+\mu_j=0$ for some $i$ and $j$, what implies in turn
that $\mu_i+\mu_j=0$. Looking at the boxed solution in (\ref{solutions}),
we see that this is indeed the case:
$\mu_1+\mu_3=0$ if $\alpha_1+\ldots +\alpha_4 = \epsilon$.
Note that {\it all} functions ${\cal Z}_{Y,Y'}$ but trivial vanish
at this point.

Conformal blocks ${\cal B}^{(k)}$ have poles at the zeroes of the Kac determinants,
which occur at
$a =\pm (s_1\epsilon_1+s_2\epsilon_2)$ with all the positive
half-integers $s_1$ and $s_2$ such that, at the level $N$, $4s_1s_2\le N$.
It is easy to see that the denominators vanish at these
points, and the functions ${\cal Z}_{Y,Y'}$ acquire the poles there
- in accordance with (\ref{BvsZ}).

\section{Conclusion}

To conclude, we explicitly checked the conjecture of \cite{agt}
for the first terms of the $x$-expansion of the four-point
conformal block.
The statement is that the universal part of conformal block,
which depends only on five dimensions (four external lines
and one intermediate) and the central charge of 2d conformal theory
can be expanded into a linear
combination of Nekrasov partition functions for the conformally invariant
4d gauge $U(2)$ theory with four fundamental multiplets.
Relation between appropriately parameterized dimensions
and parameters $(a,\mu_1,\mu_2,\mu_3,\mu_4; \epsilon=\epsilon_1+\epsilon_2)$
of Nekrasov's functions is linear: if
\be
\Delta_i = \frac{\alpha_i(\epsilon-\alpha_i)}{\epsilon_1\epsilon_2},
\ \ \ \ \ \ \ c= 1 + \frac{6\epsilon^2}{\epsilon_1\epsilon_2}
\ee
then
\be
a = \alpha - \frac{\epsilon}{2}
\ee
and $\mu$'s are given by any of the eight expressions (\ref{solutions}).
Since one can simultaneously rescale all the seven parameters $a$,
$\mu_1,\ldots,\mu_4$, $\epsilon_{1},\epsilon_2$ in the $U(2)$
Nekrasov's functions, and the common factor drops out of them in the
conformal case ($\beta\sim N_f - 2N_c= 0$), the number of free
parameters is actually {\it six} -- exactly the same as that
of $\Delta_1,\ldots,\Delta_4,\Delta$ and $c$ on
the other side of the AGT relation.

This check concerns only a small part of the AGT conjecture.
In particular, we did not touch a technically trivial,
but conceptually deep
relation between perturbative part of Nekrasov's  functions and
the structure constants of the Liouville theory, which is, perhaps,
the most beautiful part of the conjecture.
The generalizations are obvious,
however explicit checks become increasingly complicated and
- at the present level of understanding - can be performed
only by computer simulations (reported in \cite{agt}).
Still, explicit check "by hands" in the simplest case is important
to understand the statement, and our goal in this paper was
to explain how it works in some detail, avoiding yet a discussion of
underlying physics. We hope nevertheless that a performed explicit check sheds
some light to the conjectured nontrivial relation between the two-dimensional
conformal and four-dimensional gauge theory, and we are going to return to
different aspects of this relation elsewhere.

\section*{Acknowledgements}

We would like to thank A.Belavin for the illuminating discussions.

Our work was partly supported by Russian Federal Nuclear Energy
Agency and by the joint grants 09-02-90493-Ukr,
09-02-93105-CNRSL, 09-01-92440-CE. The work of A.Mar. was also supported by
Russian President's Grants of
Support for the Scientific Schools NSh-1615.2008.2, by the RFBR grant 08-01-00667,
and by the Dynasty Foundation. The work of A.Mir. was partly supported by the RFBR grant
07-02-00878, while the work of A.Mor. by the RFBR grant 07-02-00645; the work of A.Mir.
and A.Mor. was also supported by Russian President's Grants of
Support for the Scientific Schools NSh-3035.2008.2 and by the joint program
09-02-91005-ANF. A.Mar. wishes to thank Theory Division of CERN, where this work has been
completed, for the warm hospitality.

\end{document}